\documentclass[12pt]{article}
\usepackage{graphicx}
\usepackage{dcolumn}
\usepackage{bm}
\usepackage{graphics}
\usepackage{amsmath}
\usepackage{amssymb}
\usepackage{amscd}
\usepackage{afterpage}
\usepackage{float,times}
\usepackage{subfigure}
\usepackage{rotating}
\usepackage{multirow}
\usepackage{epsfig}
\usepackage{theorem}
\usepackage{moreverb}
\usepackage{euscript}
\usepackage{psfrag}

\begin{document}

\begin{center}
{\hbox to\hsize{\hfill November 2008 }}

\bigskip

\vspace{6\baselineskip}

{\Large \bf

Imprints of microcausality violation  \\ on the cosmic microwave background \\}

\bigskip

\bigskip

\bigskip

{\bf Archil Kobakhidze  \\}

\smallskip

{ \small \it
School of Physics, The University of Melbourne, Victoria 3010, Australia \\
E-mail: archilk@unimelb.edu.au
\\}

\vspace*{1.0cm}

{\bf Abstract}\\
\end{center}
\noindent
{\small We consider a modification of the Heisenberg algebra with the non-vanishing commutator of  scalar field operators. We then identify the scalar field with the second quantized inflaton fluctuation and calculate effects of microcausality violation on the temperature anisotropy of cosmic microwave background radiation. } 

\bigskip

\bigskip

\baselineskip=16pt

In the local relativistic invariant quantum field theory (in flat Minkowski space-time) microcausality is encoded in the equal time commutator,
\begin{equation}
[\phi(\vec x, t ), \phi(\vec y, t )]=0~,
\label{1}
\end{equation} 
of a Heisenberg field operator $\phi (x)$. The non-vanishing right hand side of (\ref{1}), besides the violation of microcausality, might cause violation of locality and relativistic invariance. In curved space-time the situation is more complicated. The background metric breaks relativistic invariance and also the notion of locality is lost. Nevertheless, microcausality can still hold \cite{Dubovsky:2007ac}. Heuristically, this can be understood by recalling the basic property of the (pseudo)Riemannian geometry. In the small vicinity of any point of a curved manifold the space-time can be treated as being approximately Minkowskian, and thus the condition (\ref{1}) applies at small scales. 
However, one can envisage the situation when the notions of microcausality, locality and relativistic invariance hold only approximately at large distances, and are violated at small distances. Whether the above notions are indeed the basic properties of nature must be ultimately verified by experiments. In principle, miniscule quantum effects can be detected in astronomical observations on temperature anisotropies of the cosmic microwave background (CMB) radiation, providing our universe has undergone a period of rapid inflationary expansion at early stages of its evolution. During the inflationary era  large-amplitude, small-scale quantum fluctuations of an inflaton field are stretched over the large macroscopic distances due to the rapid expansion. These quantum fluctuations are imprinted in the CMB. Therefore, one can hope to extract valuable information on quantum dynamics of the inflaton field by studying CMB temperature anisotropies. 

In this paper we would like to consider non-trivial commutator of fields (and thus violation of microcausality),
\begin{equation}
[\phi(\vec x, \eta), \phi(\vec{y}, \eta )]=\frac{i}{a^{2}(\eta)}\mu^2f (\vec x-\vec y)~,
\label{2}
\end{equation} 
in expanding conformally flat space-time with the metric, 
\begin{equation}
ds^2=a^2(\eta)(d\eta^2-d\vec x^2)~.
\label{3}
\end{equation}
In (\ref{2}) $f(\vec x- \vec y)$ is an odd function of its argument, $f(\vec x- \vec y)=-f(\vec y - \vec x)$\footnote{The dependence of $f$ on $\vec x -\vec y$ is because we have assumed that spatial translations are preserved.}, and $\mu$ is some parameter with a dimension ${\rm dim}[\phi]-{\rm dim}[f]/2$. Since $f(\vec x- \vec y)$ is the odd function, the commutator (\ref{2}) does not preserve rotational invariance. Apart from the modified commutator in (\ref{3}), we also have the standard commutators, 
\begin{equation}
[\phi(\vec x, \eta), \pi(\vec y , \eta )]=\frac{i}{a(\eta)}\delta(\vec x - \vec y)~, 
\label{4}
\end{equation}
and,
\begin{equation}
[\pi(\vec x, \eta), \pi(\vec y, \eta )]=0~,
\label{5}
\end{equation} 
where $\pi(x)$ is the operator of canonical momentum. 
Now observe that the above pair of operators $(\phi(x), \pi(x))$ with the deformed Heisenberg algebra (\ref{3}), (\ref{4}), (\ref{5}), can be mapped onto the canonical pair $(\psi(x), \pi(x))$ with the ordinary Heisenberg algebra, through the following relation, 
\begin{equation}
\phi(x)=\psi(x)-\frac{\mu^2}{2}\int f(\vec x -\vec \xi)\pi(\vec \xi, \eta)d^3\vec\xi~.
\label{6}
\end{equation}
Note that when considering ($\phi$, $\pi$) as a canonical pair, the usual Lagrangian gets modified correspondingly. We do not need the exact form of this new Lagrangian here, because all correlators can be expressed through the correlators of the standard canonical pair ($\psi$, $\pi$) through the mapping (\ref{6}). We also assume that the classical background solution holds in such a new theory. 
We identify $\phi(\vec x, \eta)$ with the second-quantized fluctuation of an inflaton field and are interested in computing two-point function $\langle 0| \phi(\vec x, \eta)\phi(\vec y, \eta)|0\rangle$ using the map (\ref{6}). The power spectrum corresponding to this correlator essentially describes cosmological perturbations which are observed through the CMB temperature fluctuations \cite{Mukhanov:2007zz}. Since the field $\psi$ is the standard quantum inflaton field, we have,
\begin{equation}
\psi(x)=\int\left[u_k(\eta){\rm e}^{i\vec k\cdot \vec x}a_k +u^{*}_k(\eta){\rm e}^{-i\vec k\cdot \vec x}a_k^{\dagger}\right]\frac{d^3\vec k}{a(\eta)(2\pi)^{3}}~,
\label{8}
\end{equation}
and 
\begin{equation}
\pi(x)=\frac{\partial (a\psi)}{\partial \eta}\equiv (a\psi)'~,
\label{9}
\end{equation}
where $a_k$ ($a_k^{~\dagger}$) are the annihilation (creation) operators,
\begin{equation}
[a_k, a_p^{~\dagger}]=(2\pi)^3\delta(\vec k-\vec p)~,~~a_{k}|0\rangle =0~.
\label{10}
\end{equation}
The mode function $u_{k}(\eta)$ satisfies the equation of motion, 
\begin{equation}
u_k''+\omega_k^2 u_k=0,~~\omega_k^2=k^2-\frac{a''}{a}~,
\label{10a}
\end{equation}
where the scale factor during the inflationary era $a(\eta)=-1/\eta H$ (and thus, $a''/a\approx 2/\eta^2$) and the expansion rate $H$ is (nearly) $\eta$-independent. With the initial conditions taken as $u_k(\eta_i)=1/\sqrt{2\omega_k(\eta_i)}\approx 1/\sqrt{2k}$ and $u'_k(\eta_i)=i\sqrt{\omega_k(\eta_i)/2}\approx i\sqrt{k/2}$ the solution of eq. (\ref{10a}) for the modes of astrophysical interest within the horizon, $k>>|aH|\approx 1/|\eta|$ ) reads,
\begin{equation}
u_k(\eta)=\frac{e^{-ik(\eta-\eta_i)}}{\sqrt{2k}}\left(\frac{i}{k\eta}-1\right)~,
\label{10b}
\end{equation}
and, correspondingly,
\begin{equation}
u'_k(\eta)=\frac{ike^{-ik(\eta-\eta_i)}}{\sqrt{2k}}\left(1-\frac{i}{k\eta}-\frac{1}{k^2\eta^2}\right)~.
\label{10c}
\end{equation}
Now, using the relation (\ref{6}), we obtain for the non-standard quantum inflaton fluctuation, 
\begin{eqnarray}
\phi(x)=\int\left[\left(u_k(\eta)-\frac{\mu^2}{2}(2\pi)^{3/2}F(\vec k)u'_k(\eta)\right){\rm e}^{i\vec k\cdot \vec x}a_k \right.\nonumber \\
+\left. \left(u^{*}_k(\eta) -\frac{\mu^2}{2}(2\pi)^{3/2}F(\vec k)u^{*'}_k(\eta)\right){\rm e}^{-i\vec k\cdot \vec x}a^{~\dagger}_k \right]\frac{d^3\vec k}{a(\eta)(2\pi)^{3}}~,
\label{11}
\end{eqnarray} 
where $F(\vec k)$ is the Fourier image of $f(\vec x)$, $F(\vec k)=\int\frac{d^3\vec k}{(2\pi)^{3/2}}f(\vec x){\rm e}^{-i\vec k\cdot \vec x}$. The $\phi\phi$ two point function then takes the form:
\begin{equation}
\langle 0| \phi(\vec x, \eta)\phi(\vec y, \eta)|0\rangle = \int \frac{1}{a^2(\eta)}\left| u_k(\eta) -\frac{\mu^2}{2}(2\pi)^{3/2}F(\vec k)u'_k(\eta)\right|^2\frac{\sin(kr)}{r}kdk~,
\label{12}
\end{equation}
where $r\equiv |\vec x-\vec y|$. From the above eq.(\ref{12}) we extract the power spectrum:
\begin{eqnarray}
P_{\phi}(\vec k)=\frac{1}{a^2(\eta)}\left| u_k(\eta) -\frac{\mu^2}{2}(2\pi)^{3/2}F(\vec k)u'_k(\eta)\right|^2
\label{13}
\end{eqnarray} 
Note that the power spectrum depends on the vector $\vec k$ through $F(\vec k)$, which indicates that the rotational invariance is broken alongside with the violation of microcausality\footnote{Recently, there were a number of studies of spatially anisotropic CMB temperature fluctuations, motivated by an anisotropic inflation involving background vector \cite{Gumrukcuoglu:2006xj}, \cite{Ackerman:2007nb} (see also \cite{Donoghue:2007ze}) or spinor \cite{Boehmer:2007ut} fields, anisotropic dark energy \cite{Koivisto:2008ig}, cosmological magnetic field \cite{Kahniashvili:2008sh} or space-time noncommutativity \cite{Akofor:2007fv}. In the later case causality is also violated.}. The ordinary rotationally invariant spectrum $P_{\phi}(k)=\frac{\left| u_k(\eta)\right|^2}{a^2(\eta)}$ is obtained from (\ref{13}) in the limit $\mu \to 0$. 
Taking the limit $|k\eta| <<1|$ (which is the case at the end of inflation) in (\ref{13}) and evaluating $P_{\phi}(\vec k)$ for the mode $k$ which crosses the horizon, i.e. when $a(\eta)H=k$, say for $\eta=\eta_0$, we compute the power spectrum of the scalar metric perturbations (in the conformal Newtonian gauge):
\begin{eqnarray}
P_{\Phi}(\vec k)=\frac{16\pi G_N}{9\epsilon}\left.P_{\phi}(\vec k)\right|_{k=a(\eta_0)H}\approx\left.
\frac{16\pi G_N}{9\epsilon}\frac{H^2}{2k^3}\left |1+\mu^2\pi^3{\rm e}^{-i\frac{\pi}{4}}F(\vec k)k^2\right|^2\right |_{k=a(\eta_0)H}~,
\label{14}
\end{eqnarray}
where $G_N$ is the Newton's constant and $\epsilon$ is the slow-roll parameter. Finally, given the above power spectrum, we determine statistical properties of the CMB temperature fluctuations $\frac{\delta T}{T}(\vec n)$ in a given direction of the unit vector $\vec n$. It is customary to decompose the temperature fluctuations that we observe in spherical harmonics, 
\begin{equation}
\frac{\delta T}{T}(\vec n)=\sum_{lm}a_{lm}Y_{lm}(\vec n)~.
\label{17}
\end{equation}
The statistical properties of the temperature fluctuations are encoded in the second order  correlators, which are expressed through the primordial power spectrum (\ref{14}) and radiation transfer functions $\Delta_l (k)$\footnote{Here we assume that in the regime where the transfer functions are effective, the effects of microcausality and rotational symmetry violation are negligible. Therefore $\Delta_l (k)$ depends on modulus $k$ only.} as,
\begin{eqnarray}
C_{ll'mm'}\equiv \langle a_{lm}a^{*}_{l'm'} \rangle =
C_{ll'mm'}^{(0)}+C_{ll'mm'}^{(1)}+C_{ll'mm'}^{(2)}~,
\label{18}
\end{eqnarray}
where, 
\begin{equation}
C_{ll'mm'}=\frac{16G_N H^2}{9\epsilon}\delta_{ll'}\delta_{mm'}\int \Delta_l^2(k)\frac{dk}{k}~,
\label{19}
\end{equation}
is the ordinary isotropic correlator of the scale-invariant perturbations, and the last two terms are anisotropic contributions, coming from the violation of microcausality/rotational invariance, 
\begin{equation}
C_{ll'mm'}^{(1)}=\frac{16G_N H^2}{9\epsilon}\pi^3\mu^2(-i)^{l-l'}\int \Delta_l(k)\Delta_{l'}(k){\rm Re}\left({\rm e}^{-i\frac{\pi}{4}}F(\vec k)\right)Y^*_{lm}(\vec k/k)Y_{l'm'}(\vec k/k)\frac{d^3{\vec k}}{k}~,
\label{19}
\end{equation}
and
\begin{equation}
C_{ll'mm'}^{(2)}=\frac{16G_N H^2}{9\epsilon}\pi^5\mu^4(-i)^{l-l'}\int \Delta_l(k)\Delta_{l'}(k)\left|F(\vec k)\right|^2Y^*_{lm}(\vec k/k)Y_{l'm'}(\vec k/k)k d^3{\vec k}~,
\label{20}
\end{equation}
The eqs. (\ref{19}) and (\ref{20}) represent generic corrections of the order $\mu^2$ and $\mu^4$, respectively. Both of this contributions are determined through the \emph{a priory} unknown function $F(\vec k)$. Obviously, from the phenomenological point of view, there are many choices for this function and different choices will lead to different results. As an explicit example (and for the sake of simplicity), we consider here $f(\vec x-\vec y)=\delta(x^1-y^1)\delta(x^2-y^2)\delta'(x^3-y^3)$ in a coordinate system of the CMB observation. With this choice we have a residual $x^1-x^2$ planar symmetry. The Fourier transform reads\footnote{If we take, e.g. $f(\vec x-\vec y)=\delta (x^1-y^1)\delta (x^2-y^2){\rm sgn}(x^3-y^3)$ instead, we obtain 
$F(\vec k)=\frac{i}{2^{1/2}\pi^{3/2}}\frac{1}{k_3}$. In this case the effects are enhanced at large scales, i.e. for modes with small $k_3$. It would be interesting to see, whether such infrared effects could mimic late time (anisotropic) acceleration of the universe through the backreaction on the background metric.}: 
\begin{equation}
F(\vec k)=\frac{-i}{(2\pi)^{3/2}}k\cos\theta~.
\label{21}
\end{equation}
Substituting (\ref{21}) into (\ref{19}) and performing angular integration, we obtain for the anisotropic parity-odd contribution, 
\begin{eqnarray}
C^{(1)}_{ll'mm'}=\frac{16G_N H^2}{9\epsilon}\frac{i\pi^{3/2}\mu^2}{4}\xi_{ll'mm'}\int \Delta_l(k)\Delta_{l'}(k)k^2dk~, \nonumber \\ \nonumber \\
\xi_{ll'mm'}=
\delta_{mm'}\left[\sqrt{\frac{(l+1)^2-m^2}{(2l+1)(2l+3)}}\delta_{(l+1)l'}
+\sqrt{\frac{l^2-m^2}{4l^2-1}}\delta_{(l-1)l'}\right]~.
\label{22}
\end{eqnarray} 
Similar computations give for the parity-even contribution (\ref{20}):
\begin{eqnarray}
C^{(2)}_{ll'mm'}=\frac{-16G_N H^2}{9\epsilon}\frac{\pi^{2}\mu^4}{2}\chi_{ll'mm'}\int \Delta_l(k)\Delta_{l'}(k)k^4dk~, \nonumber \\ \nonumber \\
\chi_{ll'mm'}=
\delta_{mm'}\left[\left(\frac{((l+1)^2-m^2)}{(2l+1)(2l+3)}+\frac{(l^2-m^2)}{(2l-1)(2l+1)}\right)\delta_{ll'} \right.\nonumber \\
+\left.\sqrt{\frac{((l+1)^2-m^2)((l+2)^2-m^2)}{(2l+1)(2l+3)^2(2l+5)}}\delta_{(l+2)l'}
+\sqrt{\frac{(l^2-m^2)((l-1)^2-m^2)}{(2l-1)(2l+1)^2(2l+3)}}\delta_{(l-2)l'}\right]~.
\label{23}
\end{eqnarray}
In conclusion, we have considered the effects of microcausality violation in quantum inflaton field on the CMB. Microcausality violation described by the non-standard commutator (\ref{2}) is supplemented by the violation of the rotational invariance. As a result, the ordinary isotropic and (nearly) scale-invariant primordial perturbations receive anisotropic and scale non-invariant corrections. These corrections modify statistical properties of the CMB temperature fluctuations as described by eqs. (\ref{22}) and (\ref{23}) for the explicit case of the microcausality violation considered in this paper.

\subparagraph{Acknowledgments.}

This work was supported by the Australian Research Council.

\end{document}